\documentclass[aps, prl, twocolumn,showpacs,tightenline]{revtex4}

\usepackage{epsfig}
\begin{document}

\title{Photon-bunching measurement after 2$\times$25\,km of standard optical fibers}
\author{M. Halder} \email{matthaeus.halder@physics.unige.ch}
\author{S. Tanzilli}
\author{H. de Riedmatten}
\author{A. Beveratos}
\author{H. Zbinden}
\author{N.~Gisin}
\affiliation{Group of Applied Physics, University of Geneva, 1211
Geneva 4, Switzerland}

\date{\today}

\begin{abstract}
To show the feasibility of a long distance partial Bell-State
measurement, a Hong-Ou-Mandel experiment with coherent photons is
reported. Pairs of degenerate photons at telecom wavelength are
created by parametric down conversion in a periodically poled
lithium niobate waveguide. The photon pairs are separated in a
beam-splitter and transmitted via two fibers of 25\,km. The
wave-packets are relatively delayed and recombined on a second
beam-splitter, forming a large Mach-Zehnder interferometer.
Coincidence counts between the photons at the two output modes are
registered. The main challenge consists in the trade-off between
low count rates due to narrow filtering and length fluctuations of
the 25\,km long arms during the measurement. For balanced paths a
Hong-Ou-Mandel dip with a net visibility of 47.3 \% is observed,
which is close to the maximal theoretical value of 50\% developed
here. This proves the practicability of a long distance Bell state
measurement with two independent sources, as e.g. required in an
entanglement swapping configuration in the scale of tens of km.
\end{abstract}

\maketitle

\section{1. Introduction}

Quantum communication via telecom fibers is still limited to
around a hundred of kilometers due to fiber losses and noisy
detectors. One way to overcome this limit is the use of
entanglement swapping \cite{zuk93}, e.g. in a quantum relay
configuration as proposed in \cite{franson,yamamoto}. Furthermore
entanglement swapping is a beautiful manifestation of the oddness
of quantum mechanics and deserves thus by itself to be
demonstrated over large distances outside the lab. At the heart of
the entanglement swapping scheme lies a Bell state measurement
(BSM) \cite{weinfurter} between two photons, originating from
independent sources of entangled photon pairs. By projecting these
two independent photons onto one of the four Bell states, the two
remaining photons, formerly independent, become entangled, even
though they never interacted \cite{zuk93,zeilinger,dan}.

A complete BSM requires huge non-linearities \cite{luet}, which
are exceedingly difficult to achieve at the single photon level.
Nevertheless, a partial BSM can be realized with a simple
beam-splitter (BS). To obtain a successful BSM, the two photons
involved must be indistinguishable, i.e. must be in the same
spatial, temporal, spectral and polarization mode in order to
bunch at the BS. The experimental feasibility of a partial BSM has
already been demonstrated for polarization qubits
\cite{Bouwmeester97} as well as for time-bin qubits \cite{iwan2},
but never over tens of kilometers as required for communication
under realistic circumstances. However, a realization with distant
sources implies additional experimental difficulties, in
particular, if the photons are transmitted through optical fibers.
A recent experiment \cite{3x2} showed, that the major problem is
to maintain temporal indistinguishability, because of thermally
induced fiber length fluctuations and thus differences in the
travelling times. One possibility to relax the length stability
requirements is to use photons with a long coherence length.
However this requires narrow filters which reduce the production
rate of photon pairs. Hence longer acquisition times are necessary
and the constraints for stability become more severe.

Pairs of entangled photons can be obtained by spontaneous
parametric down-conversion (SPDC) in a $\mathcal{X}^{(2)}$
nonlinear crystal. In this process a pump photon can be converted
with a small conversion efficiency, into two entangled photons,
according to conservation of energy and momentum. However, photons
produced by SPDC show generally a large bandwidth, corresponding
to a short coherence length. Different approaches have been
considered to increase the coherence length of photons obtained by
such a spontaneous process. One possibility is to place the
nonlinear crystal in a cavity \cite{ou}. Another possibility is to
use strongly non-degenerate photons created in long crystals.
Bandwidths of 60\,GHz have recently been reported for photon pairs
at 800 and 1600\,nm \cite{mit}. But this solution is not suitable
if both photons are wanted to be at telecommunication wavelengths.
Finally a last and simple approach that we adopt here consists in
filtering a small range out of the broad emitted spectrum. Note
that this demands a high yield of the photon source, in order to
compensate the associated losses, as mentioned above.

In this paper, we present an experiment proving the feasibility of
a partial BSM over two times 25\,km of optical fibers. We create
two degenerate photons at 1550\,nm by SPDC in a periodically poled
lithium niobate (PPLN) waveguide. In order to increase their
coherence length, their initial spectral bandwidth of 80\,nm is
filtered down to 0.8\,nm (100\,GHz), corresponding to a fourier
limited coherence length of 1.3\,mm FWHM in air. As previously
shown, this kind of waveguide features a high conversion
efficiency \cite{seb}. The two photons are sent to a 25\,km long
balanced Mach-Zehnder interferometer and the necessary
indistinguishability is verified by performing a Hong-Ou-Mandel
(HOM) experiment \cite{hom}.

This experiment represents a first step towards the realization of
entanglement swapping with sources, which are separated by a long
distance. In section\,2 we calculate the theoretical predictions
for our particular system, showing that beside the HOM
interference, a conjunction of one and two photon interference is
expected. In section\,3, we present the experiment and analyze the
results, followed by a discussion on the dispersion cancellation
effect in section\,4. A conclusion is given in section\,5.

\section{2. Theoretical expectations}

To prove indistinguishability between two photons after travelling
25\,km each, we perform a HOM experiment. Thereby the two
indistinguishable photons, entering a BS in two different input
modes, are superposed. Due to their bosonic nature, they bunch
together and exit always in the same output-mode. The coincidence
rate between two detectors, one at each output of the BS, is hence
dropping to zero. These second order quantum interferences only
appear if the two photons are completely indistinguishable in
their polarization, spectral, temporal and spatial modes.
 By shifting now the relative arrival time of one
photon in respect to the other, this drop in the coincidence rate
can be observed as a dip within their overlapping range, the
so-called "HOM-dip".

 \begin{figure}[h!]
\begin{center}
\epsfig{figure=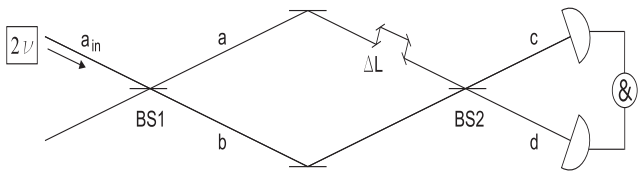,width=84mm,height=22mm}
\caption{\textit{Schematic of the experimental setup. A pair of
photons emitted by one source is split by a first beam-splitter
(BS1). These photons are passed through a Mach-Zehnder
interferometer of 2$\times$25\,km and are recombined on a second
beam-splitter (BS2). Coincidence counts between the two exit modes
are regarded, while the length of one arm is changed.}}
\label{manipdip2}
\end{center}
\end{figure}

While a complete demonstration of a partial BSM for independent
photons is usually realized with independent sources and a pulsed
pump beam \cite{zukowski95}, this is not needed here. Since we
want to check the temporal indistinguishability due to the
transmission in fibers, it suffices to take one source creating
pairs of photons, split them, and recombine them, after each has
travelled in a different fiber. A CW laser is therefore
sufficient.
  As depicted in Fig.\,1, the two photons to be overlapped emerge from
the same source completely indistinguishable, and thus only can be
separated probabilistically, e.g. by a 50/50 beam-splitter (BS1).
The two outputs of BS1 are connected to 25\,km of standard optical
fiber and recombined at BS2, which forms a balanced Mach-Zehnder
interferometer. An important consequence therefore is that both
single-photon and two-photon interferences are expected as
discussed below.

We label $a_{in}^{\dagger}$ the creation operator of input of BS1,
evolving as \emph{$a_{in}^{\dagger}\rightarrow
\frac{1}{\sqrt{2}}(ia^{\dagger}+b^{\dagger})$}, $a^{\dagger}$ and
$b^{\dagger}$ representing the upper and lower path of the
interferometer. The evolution of the phase shift is described as
\emph{$a^{\dagger}\rightarrow a^{\dagger}e^{-i\omega\tau}$}, with
$\tau=\Delta L/c$. Suppose that we have a two photon state at
input $a_{in}$ of BS1, described by
$|\psi_{in}\rangle=\int_{-\infty}^{\infty}d\omega
d\omega'G(\omega, \omega')
a_{\omega}^{\dagger}a_{\omega'}^{\dagger} |0\rangle$.
$a_{\omega}^{\dagger}$ and $a_{\omega'}^{\dagger}$ stand for the
creation operators of respectively one photon at frequency
$\omega$ and one at $\omega'$, and $G(\omega, \omega')$
characterizes the spectral distribution of the created photons.

The evolution through the two beam splitters together with the
phase shift
results in a final state, which is described by

{\setlength\arraycolsep{1pt}
\begin{center}
\begin{eqnarray}
|\psi_{in}\rangle & = &
 \int_{-\infty}^{\infty}d\omega d\omega'
G(\omega,
\omega')\nonumber\\
& &\Big(c_{\omega}^{\dagger}d_{\omega'}^{\dagger}\sin
\omega\frac{\tau}{2}\cos \omega'\frac{\tau}{2} +
d_{\omega}^{\dagger}c_{\omega'}^{\dagger}\cos
\omega\frac{\tau}{2}\sin \omega'\frac{\tau}{2}\nonumber\\
& &+ c_{\omega}^{\dagger}c_{\omega'}^{\dagger}\sin
\omega\frac{\tau}{2}\sin \omega'\frac{\tau}{2}
+d_{\omega}^{\dagger}d_{\omega'}^{\dagger}\cos
\omega\frac{\tau}{2}\cos
\tiny{\omega'\frac{\tau}{2}}\Big)|0\rangle\nonumber
\end{eqnarray}
\end{center}}

\noindent with $c^{\dagger}$ and $d^{\dagger}$ being the creation
operators of the two output modes of the second beam-splitter
(BS2). In order to calculate the coincidence probability
$P_{coinc}$, we consider $G(\omega,\omega')$ to be gaussian and
integrate over the temporal resolution of our detectors of
the order of 1\,ns. Taking into account only the terms containing
$c^{\dagger}d^{\dagger}$ and $d^{\dagger}c^{\dagger}$, $P_{coinc}$
is therefore calculated to be proportional to

\begin{displaymath}
P_{coinc} \propto
2-e^{-\frac{\tau^2}{\delta^2}}-\cos{\tau\omega_p}
\end{displaymath}

\noindent where $\omega_p$ is the pump frequency and $\delta$ the
bandwidth of the down converted photons. Note that this
calculation only holds true for a temporal path difference $\tau$
much smaller than the time resolution of the detectors.

The calculated probability $P_{coinc}$ is plotted in Fig.~2. The
form can be explained as a superposition of Franson-type
$(1-\cos{\tau\omega_p})$ 2-photon interferences \cite{franson2} and
a Hong-Ou-Mandel dip $(1-e^{-\tau^2/\delta^2})$ \cite{ou2}.
\begin{figure}[h!!]
\begin{center}
\epsfig{figure=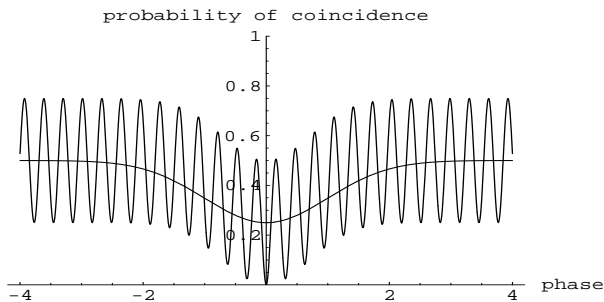,width=83mm,height=40mm}
\caption {\textit{Theoretically expected: Second order
interferences in the coincidence count rate versus the phase. The
solid line represents the count rate averaged over several phases.
Note that the phase units are arbitrary and the fringe scale has
been modified for the reasons of clearness. In reality they are
much closer.}} \label{theoriemoyen}
\end{center}
\end{figure}

The temperature fluctuations are too large to resolve the
Franson-type interferences, which are on a $\mu m$-scale, but
small enough to observe the Hong-Ou-Mandel dip (mm scale). Hence
the maximally attainable visibility of the average count-rate in
this measurement is 50\%.

The width of the dip corresponds to the convolution of the two
wave-packets. The solid line represents the average probability of
coincidences as it will be measured.

\section{3. The experiment}

\begin{figure}[h!]
\begin{center}
  \epsfig{figure=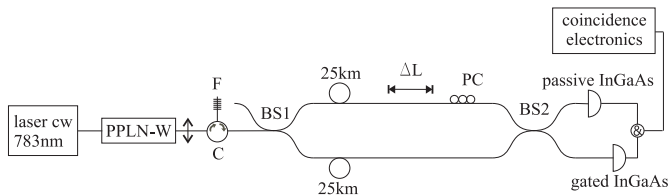,width=88mm,height=25mm}
\caption {\textit{Experimental realization of a long distance
photon bunching measurement over 2$\times$25km.A PPLN waveguide
(PPLN-W), Bragg~filter (F) coupled to a 3-port-circulator (C),
adjustable path length $\Delta L$, polarization control (PC),
beam-splitter (BS).}} \label{mandeldip}
\end{center}
\end{figure}

The experimental setup is shown in Fig.\,\ref{mandeldip}. To
create the two photons to be superposed, we pump a PPLN waveguide
\cite{hcp} with 35\,$\mu$W of a CW diode laser at 783\,nm. Pairs
of collinear energy-time entangled photons are created by SPDC
\cite{seb} with a conversion efficiency of the order of $10^{-6}$.
The waveguide is temperature stabilized at the degeneracy point,
where signal and idler are emitted at 1566\,nm and coupled into a
standard fiber with an efficiency of 18\% for each photon. This
coupling efficiency is essentially limited by the mismatch between
the guided modes in the waveguide and the collecting fiber.
Filtering is achieved by a fiber Bragg grating with 0.8\,nm FWHM.
This kind of filter reflects more than 99\% of the light in the
chosen spectral window. A circulator is used to recover the
filtered reflected light.

The pump light is sufficiently blocked by the Bragg grating and
the circulator, so no further absorbing filters are required.
Signal and idler photons are split up probabilistically by a 50/50
beam-splitter (BS1). These two output modes travel through a
Mach-Zehnder-interferometer configuration, with each arm 25.3\,km
long.

Note that the most demanding point is the stability of the length
difference between the two arms. It needs to be kept within the
coherence length of the photons during the measurement without any
active stabilization, as this is impractical in field
applications. The two fibers are on separate spools, thus a local
change in temperature affects the whole fiber length. This is not
the case in telecom fiber networks, where a local change in
temperature only concerns a certain part of the fibers. Hence we
can expect to have less fluctuations there, as already measured in
former experiments \cite{tittel} to be of the order of a few mm
per day for a 10\,km link. Even though the two fibers should
behave the same way for temperature fluctuations affecting the two
spools, we figured out a stability within 0.1\,K/h to be
necessary. This is reached by protecting them from air draught in
the laboratory and close to the conditions in the field.

To observe a HOM-dip, the arrival time of one photon at BS2 is
scanned with respect to the other one. This delay is provided by
slightly lengthening one arm mechanically. A maximal length
difference of 11\,mm is covered by a computer controlled step
motor device.

In order to insure indistinguishability at the beam-splitter
(BS2), a fiber optical polarization controller (PC) matches the
polarization modes of the two arms. At the two outputs of the
beam-splitter, photons are detected by single photon InGaAs
avalanche photodiodes (APD) which are Peltier cooled to
-30$^{o}C$. One of them is operated in passive quenching mode
\cite{john} with an efficiency of 7\% and a dark count rate of
2\,kHz. This detector triggers the second one which is optically
delayed and operated in gated mode with an efficiency of 8\% and a
dark count probability of around $10^{-5}$ per ns. Coincidences
between these two detectors are registered by a time-to-digital
converter (TDC), where only a 2\,ns time window is regarded.

 \begin{figure}[h!]
\begin{center}
\epsfig{figure=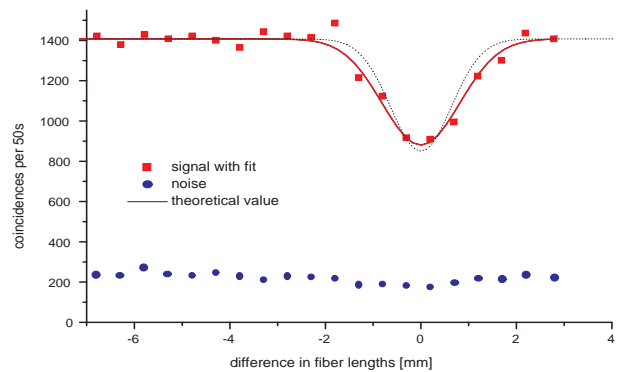,width=90mm,height=57mm}
\caption {\textit{Mandel-dip after 2$\times$25\,km: Experimental
results and fitted curve (solid line) with a maximal achievable
visibility of 50\%. The theoretical curve is represented by the
dotted line. Accidental events are plotted below.}} \label{dip5}
\end{center}
\end{figure}

In Fig.\,4, the obtained coincidence count rate is plotted as a
function of the difference between the two arms of the
Mach-Zehnder interferometer $\Delta l$. The square points show the
raw data, and the lower graph the false detection, due to dark
counts and accidental coincidences. The black line represents a
fit through these points supposing a gaussian shape. A dip with a
raw visibility of 37.6\% can be observed when the interferometer
is exactly balanced which yields a net visibility of 47.3\% when
the noise is subtracted. Although dark counts are an important
source of error in an entanglement swapping experiment, the use of
two pulsed photon pair sources, like intended in the future setup,
will allow one to trigger the detectors. Hence the accidental
coincidences will be reduce by at least one order of magnitude. In
this way, the important figure of merit in this experiment is the
net visibility.

The dotted line in Fig.\,4 gives the theoretical expected curve of
the convolution of two gaussian wave-packets. Their width
corresponds to the coherence length of the filtered wave-packets,
corrected by the refractive index of a stressed optical standard
fiber. In former experiments \cite{marc}, this value was found to
be 1.8. The width of the obtained dip is slightly broader than the
expected theoretical value.

Unfortunately, the first order interference terms cannot be
observed experimentally because of phase shifts during the
integration time. So just the mean value over several fringes is
recorded, leading to a maximal depth of the dip of 50\%.

To resolve these fringes, the path difference would required a
stability within one wavelength during the acquisition of every
data point. But due to low count-rates, integration times of 50\,s
are necessary and temperature fluctuations during this time
prevent their resolution. Note that for a temperature length
dependence of 4\,$mm\,K^{-1}\,km^{-1}$, a stability of $10^{-5}K$
would be required to resolve these interference fringes.

\section{4. Dispersion cancellation}

As can be seen from the experimental results, even after two times
25\,km neither the visibility nor the width of the HOM-dip is
significantly altered. In this section we will discuss why
chromatic dispersion in optical fibers is cancelled out in our
experiment and what it means for further experiments. A detailed
discussion can be found in \cite{steinberg92}, where we also took
the schemes from. In the scheme below we give a graphic
explanation of the cancellation, followed by an analytical
calculation.

Wave packets of a spectral width $\Delta \lambda$=0.8\,nm
correspond to a coherence length of $\Delta t$=4.25\,ps. Even
though they are broadened to 430\,ps after 25\,km of standard
optical fiber, the form of the obtained 2-photon interference dip
still corresponds to the initial coherence length. This can be
explained by the following fact. Without loss of generality, we
consider the case with a non-zero dispersion coefficient in only
one arm.

Light passing through an interferometer is subject to chromatic
dispersion like sketched in Fig.\,5. Although our spectra are
entirely in the infrared light, we will label the two photons
"red" (r) and "blue" (b), to point out their difference in
frequency. We assume dispersion only in the upper arm A.
Interference takes place for two different paths which lead to the
same detection scheme and are therefore, even in principle,
indistinguishable.

 \begin{figure}[h!]
\begin{center}
\epsfig{figure=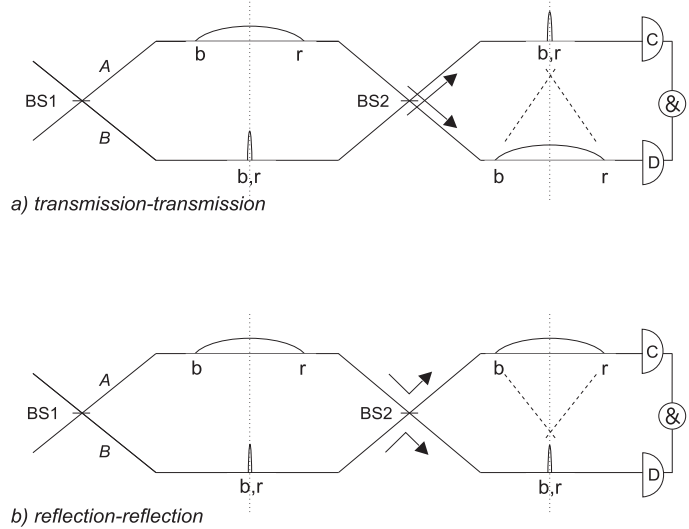,width=90mm,height=70mm}
\caption {\textit{Dispersion cancellation scheme: Even non-zero
dispersion in one arm of a Mach-Zehnder interferometer leads to
the same firing scheme of the detectors and result in an HOM-dip
with a width corresponding to the coherence time of the photons.
The dotted line represents the center of the wave-packet. BS1 and
BS2 are the two beam-splitters. It can be seen that in both cases
detector C clicks with the same time-delay after detector D. }}
\label{schemadispersion}
\end{center}
\end{figure}

The two alternatives here are either both the photons are
transmitted or reflected at BS2, shown as Fig\,5a and Fig\,5b. In
the two cases, the blue photon always arrives with the same delay
after the red one. As long as no information about the creation
time of the photon pair is available, the two cases remain
indistinguishable. But as soon as the centers of the two
wave-packets are relatively delayed by more than their coherence
length, the two possible detection events become distinguishable
and the bunching disappears. This also can be understood by
calculating the relative delay of the two photons. Be
$\tau(\omega)$ the travel time of a light pulse per length unit.
Mathematically, the effects of fiber dispersion are accounted for
by expanding $\tau(\omega)$ in a Taylor series around the center
frequency $\omega_{0}$.

\begin{eqnarray}
\tau(\omega)=\tau_{0}+\tau_{1}(\omega-\omega_{0})+\frac{1}{2}\tau_{2}(\omega-\omega_{0})^{2}+...
\label{1}
\end{eqnarray}

\noindent Note that $\tau_{0}=\frac{1}{v_{g}}$ with $v_{g}$ the
group velocity of a wave packet. The chromatic dispersion $D$
corresponds to $D~=~-\frac{\omega}{\lambda}\tau_{1}$.

Be $\omega_{0}=\frac{1}{2}\,\omega_{pump}$, and $\tau^{A}(w)$ and
$\tau^{B}(w)$ the different propagation times of fiber $A$ and
fiber $B$ at frequency~$\omega$. The time difference $\Delta\tau$
of the two cases where the blue photon ($\omega_{1}$) passes
through fiber~$A$ ($\tau^{A}(\omega_{1})$) and the red
($\omega_{2}$) through fiber~$B$ ($\tau^{B}(\omega_{2})$),
respectively vice versa ($\tau^{A}(\omega_{2}$) and
$\tau^{B}(\omega_{2})$) is described by

\begin{eqnarray}
\Delta\tau=\Big(\tau^{A}(\omega_{1})-\tau^{B}(\omega_{2})\Big)-\Big(\tau^{B}(\omega_{1})-\tau^{A}(\omega_{2})\Big)
\end{eqnarray}

By taking Eq.\,1 and Eq.\,2, $\Delta\tau$ can be calculated as

\begin{eqnarray}
\Delta\tau&=&\tau_{0}^{A}(\omega_{0})-\tau_{0}^{B}(\omega_{0})+(\tau_{1}^{A}-\tau_{1}^{B})(2\,\omega_{0}-\omega_{1}-\omega_{2})\nonumber\\
&&+\frac{1}{2}(\tau_{2}^{A}-\tau_{2}^{B})\Big((\omega_{0}-\omega_{1})^{2}+(\omega_{0}-\omega_{2})^{2})\Big)
\end{eqnarray}

The length of the two fibers can be adjusted, so that
$\tau_{0}^{A}=\tau_{0}^{B}$. For energy correlated photons
($\omega_{1}+\omega_{2}=2\,\omega_{0}$) the terms $\tau_{1}$ in
Eq.\,3 cancel out. This effect is known as dispersion
cancellation. So just the terms in $\tau_{2}$ can contribute to
$\Delta\tau\neq 0$ and cause a change of the HOM-dip. As we obtain
a net visibility of 47.3\% which is quite close to the theoretical
value of 50\%, we can conclude, that $\tau_{2}^{A}-\tau_{2}^{B}$
should be negligibly small.

For the objective of a Bell-state measurement with photons
originating from different sources, the energy-correlation is no
longer fulfilled. In this case, $\tau_{1}^{A}-\tau_{1}^{B}$ has to
be small within the coherence time and dispersion cancellation
only works for similar dispersion in both paths. For standard
telecom fibers (SMF-28) typical values of $D$ between 16.8\, and
17,9\,$ps\,nm^{-1}\,km^{-1}$ can be found. In our case, that leads
to a variation of $\tau(\omega)$ of 0.14\,$ps\,km^{-1}$ for a
bandwidth of 0.8\,nm. This means a limitation to less than 30\,km,
before this variation becomes larger than the coherence length of
the photon of 4.25\,ps. Beyond this limit, the fibers have to be
chosen to have similar chromatic dispersion or compensated
individually in order to still obtain 2-photon interferences. This
again leads to a dip width corresponding to the coherence length
of the two photons \cite{3x2}. Equal chromatic dispersion in both
arms can, in principle, be compensated by adding a dispersive
medium right in front of the detectors with an dispersion index
inverse to that of the fibers.

On the other hand, as light is polarized and needs to be in the
same mode at BS2, we have to control the polarization. Because of
the coherence length of our photons, it doesn't depolarize and
polarization mode dispersion (PMD) doesn't significantly affect
the results over this distance if high quality fibers are used.

\section{5. Conclusion}

In conclusion, we realized a proof-of-principle experiment of a
long distant HOM experiment over two times 25\,km. A SPDC source
creates signal and idler photons  at telecom wavelength with
100\,GHz bandwidth, corresponding to a coherence length of 1.3\,mm
in air. They  pass through a 25\,km Mach-Zehnder interferometer
and are recombined on a beam-splitter again. A HOM-dip over this
range has been observed, achieving a net visibility of 47.3\%,
which is quite close to the the maximal theoretically value of
50\%. This proofs the indistinguishability of photon which travel
via different fibers. It has been demonstrated that even after
2$\times 25$\,km, chromatic dispersion and PMD have no significant
impact on the obtained measurements. This represents a first step
towards long distance entanglement swapping, as required in real
world quantum communication. The following step will be a
realization with photons coming from two independent sources.

We acknowledge the support of HC Photonics for supplying the
waveguide. Financial support from the Swiss NCCR-quantum photonics
and the European IST-project RamboQ are gratefully acknowledged.
Thanks are due to J.-D.~Gautier and C.~Barreiro for technical
support and V. Scarani, B. Kraus and S. Iblisdir for theoretical
support. Finally, we thank Z.Y. Ou for pointing out a mistake in
an earlier version.


\begin{references}



\bibitem{zuk93} M. Zukowski, A. Zeilinger, M. A. Horne, and A. K. Ekert Phys. Rev. Lett. {\bf71}, 4287-4290 (1993).
\bibitem{franson} B. C. Jacobs, T. B. Pittman and J. D. Franson, Phys. Rev. A \textbf{66}, 052307 (2002).
\bibitem{yamamoto} E. Waks, A. Zeevi and Y. Yamamoto, Phys. Rev. A \textbf{65}, 052310 (2002).
\bibitem{dan} D. Collins et al, quant-ph/0311101, Accepted for publication in J. Mod. Opt

\bibitem{weinfurter} H. Weinfurter, Europhys. Lett. \textbf{25}, 559 (1994).
\bibitem{zeilinger} J.~W.~Pan, D. Bouwmeester, H.~Weinfurther, A.~Zeilinger, Phys. Rev. Lett. \textbf{80}, 18 (1998).

\bibitem{luet} N. L\"{u}tkenhaus, J. Calsamiglia and K.-A. Suominen, Phys. Rev. A \textbf{59}, 3295 (1999).
\bibitem{Bouwmeester97} D. Bouwmeester, Jian-Wei Pan, Klaus Mattle, Manfred Eibl, Harald Weinfurter and Anton Zeilinge, Nature {\bf 390}, 575, (1997).

\bibitem{iwan2} I. Marcikic, H. de Riedmatten, W. Tittel, H. Zbinden and N. Gisin,  Nature \textbf{421}, 509 (2003).
\bibitem{3x2} H. de Riedmatten, I. Marcikic, W. Tittel, H. Zbinden, D. Collins and N. Gisin, Phys. Rev. Lett., \textbf{92}, 047904 (2004).
\bibitem{ou} Z. Y. Ou and Y. J. Lu, Phys. Rev. Lett. \textbf{83}, 2556-2559 (1999).

\bibitem{mit} F. K\"{o}nig, E. J. Mason, F. N. C. Wong and M.~A.~Albota, proceedings of QIPC (2003).
\bibitem{seb} S.~Tanzilli, H. de Riedmatten, W. Tittel, H. Zbinden, P. Baldi, M. De Micheli, D. B. Ostrowsky and N. Gisin, Electr. Lett. \textbf{37}, 26 (2001).
\bibitem{hom} C. K. Hong, Z. Y. Ou and L. Mandel, Phys. Rev. Lett. \textbf{59}, 2044 (1987).

\bibitem{franson2} J. D. Franson, Phys. Rev. Lett. \textbf{62}, 2205 (1989).

\bibitem{ou2} Y. J. Lu, R. L. Campbell and Z. Y. Ou, Phys. Rev. Lett. \textbf{91}, 163602 (2003).
\bibitem{zukowski95}M. Zukowski \textit{et al}, Ann. NY Acad. Sci. \textbf{775}, 91-102 (1995).
\bibitem{hcp} HC Photonics Corp., Taiwan.

\bibitem{tittel} H. Zbinden, J. Brendel, N. Gisin and W. Tittel, Phys. Rev. A \textbf{63}, 022111 (2001).
\bibitem{john} J.~Rarity, T. E. Wall, K. D. Ridley, P.~C.~M.~Owens, P.~Tapster, Appl. Opt. \textbf{39}, 36 (2000).

\bibitem{marc} P. Oberson, B. Huttner and N. Gisin, Opt. Lett. \textbf{24}, 7 (1999).
\bibitem{steinberg92} A. M. Steinberg, P. G. Kwiat and R. Y. Chiao, Phys. Rev. A, \textbf{45}, 6656 (1992).


\end{references}
\end{document}